\pdfoutput=1

\documentclass[11pt]{article}

\usepackage{acl}

\usepackage{times}
\usepackage{latexsym}

\usepackage[T1]{fontenc}

\usepackage[utf8]{inputenc}

\usepackage{microtype}

\title{Reproducibility Signals in Science: A preliminary analysis}

\author{Akhil Pandey Akella \\
  Dept. of Computer Science \\
  Northern Illinois University \\
  \texttt{aakella@niu.edu} \\\And
  Hamed Alhoori \\
  Dept. of Computer Science \\
  Northern Illinois University \\
  \texttt{alhoori@niu.edu} \\\And
  David Koop \\
  Dept. of Computer Science \\
  Northern Illinois University \\
  \texttt{dakoop@niu.edu} \\}

\begin{document}
\maketitle
\begin{abstract}
Reproducibility is an important feature of science; experiments are retested, and analyses are repeated. Trust in the findings increases when consistent results are achieved. Despite the importance of reproducibility, significant work is often involved in these efforts, and some published findings may not be reproducible due to oversights or errors. In this paper, we examine a myriad of features in scholarly articles published in computer science conferences and journals and test how they correlate with reproducibility. We collected data from three different sources that labeled publications as either reproducible or irreproducible and employed statistical significance tests to identify features of those publications that hold clues about reproducibility. We found the readability of the scholarly article and accessibility of the software artifacts through hyperlinks to be strong signals noticeable amongst reproducible scholarly articles.
\end{abstract}

\section{Introduction}

Transparency in the scientific process accelerates scientific discovery and strengthens public opinions on scientifically driven matters. Reproducibility plays a crucial role in aiding this transparency, and it is encouraging to have a consensus in the scientific community to address the problem of reproducibility in science. Policymakers, government entities, open source communities, peer-reviewed journals, conferences, and the academic community at large have a shared responsibility to promote reproducible research. Effective dissemination of science cannot happen without trust and integrity in the scientific process. Practically, reproducible science has a first-hand impact in notable places such as research labs, classrooms, industries, and academia. Lack of reproducible research could restrict attaining a deeper understanding of the original researcher’s thought process and, therefore, severely impact people involved in the communities mentioned earlier.

The concept of reproducibility is intricate and stratified with different but complementary issues. Before we attempt to understand how to approach the problem of reproducibility, we must first provide some definition of what we mean by this term in this context. Studies such as \citep{Gundersen2018-oi, cohen-etal-2018-three, Barba2018} highlight how the definition of \textit{reproducibility} varies across different studies and disciplines and how differing definitions can result in confusion. For that reason, the flexible definition presented in \citet{Gundersen2018-oi} is appealing: ``the ability of an independent research team to produce the same results using the same method based on the documentation made by the original research team.'' Collective efforts from various players of the research community such as publishers, conference organizers, and journals in promoting good practices for ensuring reproducibility in the experimentation process is refreshing, but there is still a lack of agreement on what exactly constitutes a ``good practice'' which is a concern. 

In this study, we attempt to understand the relationship between the structure of science \cite{THELWALL2019555} and the concept of reproducibility by using statistical significance tests. In doing so, our emphasis is to examine epistemic opacity \cite{Newman2015EpistemicOC} of linguistic features and structural features concerning reproducibility. We achieve this by running numerous hypothesis tests and identifying the significant factors affecting the reproducibility of scholarly articles. Our goal is to utilize statistical tests to pick signals that could help identify articles requiring more (or less) effort to reproduce.

\section{Related Work}
Reproducibility is an important concept that affects large communities in general \cite{Mede2020, Hutson2018}. The breadth of literature on reproducibility spanning different disciplines \cite{Open_Science_Collaboration2012-fu, prinz2011believe, begley2012drug, peers2012search} has broadly focused on either performing large meta-analyses that reproduce a large set of scholarly articles or qualitative studies that encourage researchers to adopt a certain methodology.

Our study falls in line with the studies that attempts to quantify the factors important for reproducibility, e.g.~\cite{Raff2019}. Identifying such important factors would also be helpful in building machine learning models that can estimate the degree of reproducibility in scholarly articles\cite{Yang2020}.

\section{Data}

While scientific publications often follow similar structures, there is significant freedom in how ideas are communicated and expressed. This lack of rigidity allows authors to weave stories around fundamental ideas, and the absorption of particular ideas can sometimes be related to how they are presented. We  are interested in whether the structure of a publication reveals anything about its potential for (ir-)reproducibility. To examine this, we compiled a collection of scholarly articles that have been evaluated as either reproducible or irreproducible from three different sources. For each article, we gathered comprehensive metadata and extracted structural and linguistic features. These collections of articles include: 

\begin{itemize}
    \item \textbf{Brown University}: Collberg et al.~\cite{Collberg2015-xr} conducted a meta-analysis that involved steps in reproducing scholarly articles published in ACM computer science conferences and journals. They found that nearly 50 percent of the examined scholarly articles required extra effort to reproduce the articles. Computer scientists at Brown University led an effort named ``Examining Reproducibility in Computer Science" to crowdsource a reexamination of this study~\cite{ReproducibilityInCS}. They performed a meta-analysis of the original study and offered new insights. The data collected provides significant detail about the effort involved in reproducing the studies in the original publications. The current repository provides results for 207 papers; 142 are classified as reproducible and 65 as non-reproducible. 
    \item \textbf{Retraction Watch Database (RetractionDB)}: The Retraction Watch Database stores information about scholarly articles that are retracted from conferences and journals~\cite{RetractionDB}. It also logs information about the subject/area to which the scholarly article belongs, the country where the article is published, the name of the publisher, the journal name, and most importantly, the reason why the article was retracted. We used this database to find all the scholarly articles in the field of computer science that were retracted under reasons surrounding results not being reproducible, and 34 papers fit these criteria.
    \item \textbf{Badged ACM Papers}: The Association for Computing Machinery (ACM) has introduced badges as a way to signal when publications have been successfully reproduced. We began with 176 articles that were badged as having results reproduced. Of these, 90 were badged as having Reusable Artifacts, and 70 of those had a Functional Artifact badge. We were able to obtain 64 of the papers that had ``Results Reproduced'' badges and received both a Reusable Artifact and a Functional Artifact badge.
\end{itemize}

From each of the three sources, we used the available metadata to locate each article. In some cases, we searched by article and authors' names to obtain a DOI or, in some cases, a URL for an article. If we were unable to unambiguously determine this information, the article was dropped from the dataset. Using the DOI, we were able to obtain further metadata and the full text of the article, usually in PDF format. After filling out the metadata and obtaining the full text, we had 305 papers in total; 206 were classified as reproducible, and 99 were classified as non-reproducible. Data and code will be made available as supplementary information upon publishing.

\section{Methodology}
\subsection{Feature Engineering}

The motivation for considering the below features stems from the shared intuitions highlighted in \cite{Gundersen2018, Gundersen2020, Raff2019} along with checklists from popular publishing venues such as NeurIPS, ICML, etc.
		
\begin{enumerate}
    \begin{table}[htbp]
	\caption{List of Structural Features and respective Point Biserial Correlations against target variable}
	\centering
	\begin{tabular}{ p{4.25cm} p{1.75cm} }
		\hline \textbf{Feature} & \textbf{p-value} \\
        Presence of Introduction Section &0.0808 \\
        Presence of Methodology Section	&0.3112 \\
        Presence of Results Section	&0.7006 \\
        Number of Pages	&0.1630 \\
        Number of Images &0.3571 \\
        Number of Tables &0.7187 \\
        Number of Algorithms &0.0654 \\
        Number of Hyperlinks &0.0028 \\
        Number of Equations &0.4212  \\ \hline
	\end{tabular}
	\label{tab:struc-features-repr}
\end{table}
	\item \textbf{Structural features:} Quantitative and qualitative information pertaining to the structure of the scholarly article. This includes information about the existence of particular sections as well as counts of the tables, figures, or algorithms in a given scholarly article. We developed python modules to parse the PDF of the scholarly article in order to extract this information. The features along with respective Point Biserial correlations are mentioned in Table~\ref{tab:struc-features-repr}.
	
	\item \textbf{Linguistic features:} Linguistic indicators quantifying different metrics based on the language used in the scholarly article to differentiate the writing styles of various authors. These indicators include Word count, Average word length, Average sentence length, Frequency of words greater than average word length, Syllable count, and Yule’s I measure of lexical diversity \cite{yule2014statistical}. These features are general to computational linguistics and are easily understandable. Additionally, we considered metrics such as \textit{Complex words}, which refer to the number of polysyllable words in a given text. This feature was extracted using the python \textit{textblob} library. \textit{Mean Readability} was measured by obtaining the mean of readability metrics such as Flesch Reading Ease Level, SMOG Index, Coleman-Liau index, Automated Readability Index, Dale-Chall Readability Score, Linsear Write Formula, and Gunning FOG. We obtained the values from $textstat$, a python package, to obtain the readability metrics. We also collected the \textit{Sentiment} score for the full text of a given scholarly article and attached a sentiment label (positive = 1, negative = 0) for the respective articles. A similar process was used to obtain the sentiment label for the title of the article.

\begin{table}[htbp]
	\caption{List of Linguistic Features and respective Point Biserial Correlations against target variable}
	\centering
	\begin{tabular}{ p{4.25cm} p{1.75cm} }
		\hline \textbf{Feature} & \textbf{p-value} \\
		Word count	&0.5357 \\
        Average word length	&0.2379 \\
        Frequency of words greater than average word length &0.9804 \\
        Complex words &0.8394 \\
        Syllable count	&0.7467 \\
        Yule’s I measure of lexical diversity &0.1102 \\
        Mean Readability &0.0000 \\
        Article's sentiment & 0.5659 \\
		Title's sentiment  & 0.7335 \\ \hline
	\end{tabular}
	\label{tab:ling-features-repr}
\end{table}

	We gathered this information by implementing python programs that used the python libraries such as \textit{spaCy} and \textit{NLTK} to build the methods for calculating the metrics. All of these linguistic measures were based on the full text of the scholarly article. The features along with respective Point Biserial correlations, are mentioned in Table. \ref{tab:ling-features-repr}.
\end{enumerate}

\subsection{Point Biserial Correlation}
A preliminary statistical analysis of the dependent and independent variables could be performed using correlations. Since our target is a nominal variable, we could not use \emph{Pearson} correlation or \emph{Spearman} correlation as both of them presume the target variable to be continuous. The \emph{point biserial} \cite{Gupta1960} correlation matrix measures the correlation between a dichotomous target variable and continuous variables. The results in Table \ref{tab:struc-features-repr} and Table \ref{tab:ling-features-repr} are values obtained by calculating the point biserial correlation coefficient(s) and the associated p-value(s).

\subsection{Significance tests}
	The features mentioned in Tables \ref{tab:struc-features-repr} and  \ref{tab:ling-features-repr} are a combination of ordinal and nominal attributes. In order to determine the significance of the features, we had to employ different statistical significance tests such as the \textit{Mann-Whitney U} test \cite{Mann1947} and \textit{Chi-squared} test \cite{Yates1934}.
	
\section{Results}

We computed correlations and performed statistical significance tests on the combined data sources to identify features that played a significant role in indicating the reproducibility of scholarly articles. The point biserial correlations as shown in Tables \ref{tab:struc-features-repr} and \ref{tab:ling-features-repr} suggested that only \textbf{mean readability} and \textbf{number of hyperlinks} significantly correlate with reproducibility.

The results of the \textit{Mann-Whitney U} and \textit{Chi-squared} tests show that \textbf{mean readability, number of hyperlinks, number of algorithms, average word length, and yule's measure of lexical diversity} to be statistically significant features that align and signal scholarly work that is reproducible with reasonable certainty. More significantly, the readability of a scholarly article and accessibility of software artifacts, either as code repositories, psuedo-code, or algorithms, could be considered strong indicators for reproducibility. It is important to note that these signals do not quantify or assure the reproducibility of a scholarly article but rather help identify articles that require more (or less) effort to reproduce.

\begin{table}[htbp]
	\caption{Mann-Whitney U Significance test for the numerical features}
	\centering
	\begin{tabular}{ p{4.75cm} p{1.25cm} }
		\hline \textbf{Feature} & \textbf{p-value} \\
        Yule’s I measure of lexical diversity & 0.0131 \\
        Word count & 0.6547 \\
        Average word length & 0.0003 \\
        Frequency of words greater than average word length & 0.9171 \\
        Syllable count & 0.3910 \\
        Complex words & 0.9596 \\
        Mean Readability & 0.0001 \\
        Number of Images & 0.2039 \\
        Number of Tables & 0.9586 \\
        Number of Algorithms & 0.0283 \\
        Length of the paper  & 0.5039 \\
        Number of Hyperlinks & 0.0011 \\
        Number of Equations & 0.2148 \\ \hline
	\end{tabular}
	\label{tab:mannwhit-repr}
\end{table}

Our findings were backed by results from statistical experiments such as Point Biserial Correlations, Chi-squared test, and Mann-Whitney U test, and p-values (p < 0.05) served as the basis for the significance of our findings. You can obtain a copy of the datasets, experiment setup, and additional software artifacts from Github repository. \footnote{https://github.com/reproducibilityproject/reproducibilitysignals}.

	\begin{table}[htbp]
	\caption{Chi-squared  Significance test for the categorical features}
	\centering
	\begin{tabular}{ p{4.75cm} p{1.25cm} }
		\hline \textbf{Feature} & \textbf{p-value} \\
		Presence of Introduction Section & 0.1070 \\
		Presence of Methodology Section & 0.3728 \\
		Presence of Results Section & 0.8617 \\
		Article Sentiment & 0.6646 \\
		Title Sentiment & 0.8495 \\ \hline
	\end{tabular}
	\label{tab:chi2-repr}
\end{table}

\section{Discussion}

The structure of science involves a well-formed process that begins with factual and valid data, continues through detailed descriptions of experimental procedures, and follows on to clearly presented results. The scientific process has many tenets, but these represent some. They have been promulgated over the years to allow the scientific process to flourish with checks and balances in the form of peer reviews. Contextually, factors such as discipline, year, type of scientific study, etc., play a major role in identifying the effort required to reproduce articles. Therefore, the dataset we built is an essential factor to consider while interpreting our findings that the readability of the scholarly article and accessibility of the software artifacts through hyperlinks are significant features among reproducible scholarly articles. Our motivation is to discover additional latent variables that consider these contextual factors while identifying the effort required to reproduce articles. 

\section{Conclusions and Future Work}

In this study, our pursuit of identifying features that can signal reproducible science involved correlations and significance tests. We found the readability of the scholarly article and accessibility of the software artifacts through hyperlinks to be significant features among reproducible scholarly articles. Our code repository with data and experiments will be available post-publishing. 

In the future, we plan on expanding the scope of our study by 1) Gathering more Badged data from ACM; 2) Testing the validity of our findings against adversarial examples; and 3) Observing the effects of citing a reproducible article vs non-reproducible ones.

\section{Acknowledgement}
This work is supported in part by NSF Grant No. 2022443.

\bibliography{main}
\bibliographystyle{aclnatbib}

\end{document}